\documentstyle[aps,prl,12pt,amssymb]{revtex}

\begin{document}
\title{ The Kolmogorov-Sinai Entropy for Dilute Gases in Equilibrium}
\author{ H. van Beijeren[*], J. R. Dorfman[**], H. A. Posch[$\dagger$],
and Ch. Dellago[$\dagger,\ddagger$]}
\address{* Institute for Theoretical Physics, University of Utrecht,
Postbus 80006, 3508 TA, Utrecht, The Netherlands, \\ $**$ Institute for
Physical Science and Technology, and Department of Physics, University
of Maryland, College Park, MD, 20742, U.S.A. \\ $\dagger$ Institut
f\"{u}r Experimentalphysik, Universit\"{a}t Wien, Boltzmanngasse 5,
A-1090, Wien, Austria \\ $\ddagger$ Department of Chemistry, University
of California, Berkeley, CA, 94720, U.S.A.}
\date{\today}
\maketitle

\begin{abstract}
We use the kinetic theory of gases to compute the Kolmogorov-Sinai
entropy per particle for a dilute gas in equilibrium. For an
equilibrium system, the KS entropy, $h_{KS}$ is the sum of
all of the positive Lyapunov exponents characterizing the chaotic
behavior of the gas. We compute $h_{KS}/N$, where $N$ is the number of
particles in the gas. This quantity has a
density expansion of the form $h_{KS}/N = a\nu[-\ln{\tilde{n}} + b
+O(\tilde{n})]$, where $\nu$ is the single-particle collision frequency and $\tilde{n}$ is the reduced number density of the gas. The theoretical values for the coefficients $a$ and $b$ are compared with the results of
computer simulations, with excellent agreement for $a$, and less  than
satisfactory agreement for $b$. Possible reasons for this difference
in $b$ are discussed.
\end{abstract}

One of the important quantities characterizing the chaotic behavior of
a dynamical system is the Kolmogorov-Sinai (KS) entropy,
$h_{KS}$\cite{Ott,Ru-Ek}. If
the system is isolated and closed, {\it i.e.} there is no escape of
particles from the system, then according to Pesin's theorem \cite{Ru-Ek}, $h_{KS}$ is the sum of all the
positive Lyapunov exponents of the system, where the Lyapunov exponents
characterize the rate of exponential separation of the system's trajectories in
phase space. This quantity is of considerable interest because a
positive KS entropy implies that the system is chaotic, and because the KS entropy measures
the rate at which information about the phase space trajectories is
produced by the dynamics. Further, sums of Lyapunov exponents figure
prominently in the expressions for transport coefficients of fluids in
terms of dynamical quantities \cite{Gas-Nic,Do-Gas}. Recently methods based upon the kinetic
theory of gases have been applied to compute the chaotic properties of
simple systems such as hard disk and hard sphere Lorentz gases
\cite{vBD,vBDCPD,LvBD,DP}, as
well as Lorentz lattice gases \cite{EDNJ,AEvBD}. This work has shown that it is possible
to calculate theoretically Lyapunov exponents for Lorentz gases at low
density under a variety of equilibrium and non-equilibrium
situations. The theoretical results agree very well with computer
simulations for the same quantities. 
Here we show how kinetic theory methods can be extended to a
calculation of the equilibrium
KS entropy of dilute gases composed of $N$ particles interacting with
short range forces. This extension of the methods developed earlier
for Lorentz gases to 
systems of $N$ moving particles allows us to begin a
thorough exploration of the chaotic properties of systems with
large numbers of moving particles, as well as of Lorentz gases, both
in and outside equilibrium states. In subsequent publications 
Van Beijeren and 
Van Zon \cite{vBvZ} will describe methods to obtain 
the
largest 
positive Lyapunov exponent for a dilute gas in
equilibrium. 

We consider a gas composed of $N$ identical particles in
equilibrium. The particles obey classical mechanics, each have mass
$m =1$, and their positions and velocities are denoted by $\vec{r}_i,
\vec{v}_i$, respectively, with $i = 1,..., N$. The particles interact
with strongly repulsive, short range forces
with a finite
range of the
interparticle force,
$\sigma$. In the case that the particles are
hard disks or hard spheres, $\sigma$ is the diameter of each
particle. Our goal is to 
determine
the KS entropy per particle, $h_{KS}/N$, in the thermodynamic limit, assuming that the system is in thermal equilibrium at temperature $T$,
and at low densities with $n\sigma^{d}\ll1$, where $n$ is the
number density, $n = N/V$, $V$ is the volume of the system, and $d$ is
the dimension of the system, $d=2,3$. 

We consider the trajectory of our system on the constant
energy surface in the $2dN$-dimensional phase space, $\Gamma$,
starting from some initial point ${\bf{X}}_{0}=(\vec{r}_{1}(0),
\vec{v}_{1}(0), \vec{r}_{2}(0), \vec{v}_{2}(0), ..., \vec{r}_{N}(0),
\vec{v}_{N}(0))$. We then consider a bundle of trajectories that
start at infinitesimally nearby points in phase space, with each
trajectory in the bundle denoted by
${\bf{X}}(0)+\delta{\bf{X}}(0)$, for some infinitesimal
$\delta{\bf{X}}(0)=(\delta\vec{r}_{1}(0), \delta\vec{v}_{1}(0),...,\delta\vec{r}_{N}(0),\delta\vec{v}_{N}(0))$. The time rate of separation of this
trajectory bundle in the various directions perpendicular to the
direction of flow, if exponential, is characterized by a set of
non-zero Lyapunov exponents which are positive in the expanding
directions and negative in the contracting 
ones. The KS entropy
is, for our system, given by the sum of all of the
positive Lyapunov exponents, by Pesin's theorem \cite{Ott,Ru-Ek}. In
order to determine the Lyapunov
exponents and their sum, we need to follow the dynamics of the trajectory bundle over 
a 
time $t$
very large compared to some characteristic microscopic time, $t_0$, which for low
density systems can be taken to be the mean free time between
collisions for a typical particle. Recently Dellago, Posch and Hoover,
as well as Gaspard and Dorfman
have obtained the equations of motion for a trajectory bundle in phase
space for a system of hard disk or hard sphere particles\cite{Do-Gas,DPH,DP97}. 
These equations also apply to any system of particles with strong short
range interaction potentials, provided we neglect effects due to
collisions involving more than two particles at a time, or due to the
finite duration of collisions taking place in the fluid. To lowest
order in the density, these multiple collision and finite collision
time effects may be ignored
and the dynamics may be thought of as periods of free motion of the
particles punctuated by binary collisions between pairs of
particles. During the free motion, the positions and velocities of the
particles change with time according to:
\begin{eqnarray}
\vec{r}_{i}(t) & =& \vec{r}_{i}(0)+\vec{v}_{i}(0)t \\
\label{1}
\vec{v}_{i}(t) & = & \vec{v}_{i}(0) \,\,\,{\rm for}\,\,i=1,..., N.
\label{2}
\end{eqnarray}
Here the time $t=0$ represents the instant of the last binary
collision in the system, and time $t$ is some time between the last
binary collision and the next. When a collision takes place between
particles $k$ and $l$, say, there is an effectively instantaneous
change in the velocities of particles $k,l$ while the positions and
velocities of all of the other $N-2$ particles are
unaffected. Immediately after the $(k,l)$ collision, the positions and
velocities of the two particles, $k,l$ are given by
\begin{eqnarray}
\vec{v}'_{k} & = & \vec{v}_{k}+[\vec{v}_{lk}\cdot\hat{n}]\hat{n}, \\
\label{3}
\vec{v}'_{l} & = & \vec{v}_{l}-[\vec{v}_{lk}\cdot\hat{n}]\hat{n}, \\
\label{4}
\vec{r}_{l} & = & \vec{r}_{k}+\sigma\hat{n}.
\label{5}
\end{eqnarray}
Here $\hat{n}$ is a unit vector from the center of particle $k$ to the
center of particle $l$ at the point of closest approach during the
binary collision, $\vec{v}_{lk}=\vec{v}_{l}-\vec{v}_{k}$, and
$\vec{v}_{lk}\cdot\hat{n}\leq 0$. The free motion of the other $N-2$
particles $i\neq k,l$ is not affected by the $k,l$ collision. Now we
consider the other trajectories in the nearby bundle. We assume that
the bundle is sufficiently narrow that all trajectories in the bundle
exhibit the same collisions in the same sequence, with slight
differences in the times of the collisions, in the velocities and
positions before and after the collisions, and in the points of closest
approach. The analysis of Dellago, Posch, and Hoover leads immediately
to equations of motion for the deviations of the positions and
velocities in the trajectory bundle from the main trajectory. During
the free motion between collisions, $\delta\vec{r}_{i}$, and
$\delta\vec{v}_{i}$ evolve according to
\begin{eqnarray}
\delta\vec{r}_{i}(t) & = & \delta\vec{r}_{i}(0)+t\delta\vec{v}_{i}(0) \\
\label{6}
\delta\vec{v}_{i}(t) & = & \delta\vec{v}_{i}(0).
\label{7}
\end{eqnarray}
Whenever a collision takes place in the system, the deviations in
positions and velocities of the colliding particles are directly
affected, while those of the other particles are not. The velocity
deviations for the colliding pair, say particles $k,l$, can be
obtained from Eqs. (3,4) by taking linear deviations, that is,
\begin{eqnarray}
\delta\vec{V}'_{lk} & \equiv & \frac{1}{2}[\delta\vec{v}'_{k}
+\delta\vec{v}'_{l}]=\delta\vec{V}_{lk}
\\
\label{8}
\delta\vec{v}'_{lk} & \equiv & \delta\vec{v}'_{l}-\delta\vec{v}'_{k}=\delta\vec{v}_{lk}-2(\delta\vec{v}_{lk}\cdot\hat{n})\hat{n}-2[(\vec{v}_{lk}\cdot\delta\hat{n})\hat{n}+(\vec{v}_{lk}\cdot\hat{n})\delta\hat{n}],
\label{9}
\end{eqnarray}
where for later convenience we have introduced the center of mass
velocity $\vec{V}_{lk}=(1/2)[\vec{v}_{k}+\vec{v}_{l}]$, in addition to
the relative velocity of the colliding pair of particles. Here
$\delta\hat{n}$ is the infinitesimal displacement of the unit vector
in the direction of
closest approach, due to the displacement of the trajectory, and
$\hat{n}\cdot\delta\hat{n}=0$. To 
calculate
$\delta\hat{n}$ we use the
coupled equations for the location of the collision for both the
undisplaced and the displaced trajectories:
\begin{eqnarray}
\sigma
\hat{n} & = & \vec{r}_{l,0}+\tau_{l}\vec{v}_{l}-\vec{r}_{k,0}-\tau_{k}\vec{v}_{k}
\\
\label{10}
\sigma \delta\hat{n} & = &
\delta\vec{r}_{l,0}-\delta\vec{r}_{k,0}+\vec{v}_{lk}\delta\tau_{lk}+\tau_{l}\delta\vec{v}_{l}-\tau_{k}\delta\vec{v}_{k},
\label{11}
\end{eqnarray}
where we have kept terms to linear order in the deviations. Here
$\vec{r}_{l,0}$ is the position of particle $l$ at the instant of its last collision with another
particle in the system, and similarly for $\vec{r}_{k,0}$. Also
$\tau_{l}$ is the time elapsed from the last collision of particle $l$
until its collision with $k$, with a similar definition for
$\tau_{k}$. Finally $\delta\tau_{lk}$ is the difference in time between the
$l,k$ collision for the displaced and the undisplaced trajectories.
We assume here that in our calculation of $h_{KS}/N$, the dominant terms will come
from
terms proportional to $T_{kl}=
(\tau_{k}+\tau_{l})
/2$ 
which is on the
order of the free time between collisions of particles in the
system and which scales inversely with the density of the gas. For low
densities this free time can be quite large, and our assumption is that we 
can neglect the term $\delta\vec{r}_{l,0}-\delta\vec{r}_{k,0}$ in
Eq. (\ref{11}) compared to terms proportional to the times between
collisions for each of the particles.
 This assumption will be
discussed further in the concluding remarks. Then using the condition
$\hat{n}\cdot\delta\hat{n}=0$, and changing to center-of-mass and
relative velocities, we finally obtain equations for the change in the
displaced center of mass and displaced relative velocities of
particles $k,l$ at the $k,l$ collision: 
\begin{eqnarray}
 \delta \vec{v}^{\prime}_{i} & = & \delta \vec{v}_{i}
\,\,\, {\rm for} \,\,
i \neq k,l  
\\
\label{12}
\delta\vec{V}'_{kl} &  = & \delta\vec{V}_{kl} \\
\label{13}
\delta\vec{v}'_{lk} 
&  = &  \delta\vec{v}_{lk}-2(\delta\vec{v}_{lk}\cdot\hat{n})\hat{n} \nonumber\\
& & - \left( \frac{2}{\sigma} \right) \left\{ t_{lk} \left[ \left(
(\vec{v}_{lk}\cdot\delta\vec{V}_{kl})-\frac{v_{lk}^{2}}{(\vec{v}_{lk}\cdot\hat{n})}(\delta\vec{V}_{kl}\cdot\hat{n})
 \right) \hat{n}
 +\delta\vec{V}_{kl}(\vec{v}_{lk}\cdot\hat{n})-\vec{v}_{lk}(\delta\vec{V}_{kl}\cdot\hat{n})
\right]\right. \nonumber\\
& & +\left. T_{kl} \left[ \left(
(\vec{v}_{lk}\cdot\delta\vec{v}_{lk})-\frac{v_{lk}^{2}}{(\vec{v}_{lk}\cdot\hat{n})}(\delta\vec{v}_{lk}\cdot\hat{n})
\right)
\hat{n}+(\vec{v}_{lk}\cdot\hat{n})\delta\vec{v}_{lk}-\vec{v}_{lk}(\delta\vec{v}_{lk}\cdot\hat{n})
\right] \right\}.
\label{14}
\end{eqnarray}
Here $t_{lk}=\tau_{l}-\tau_{k}$. Using these equations we can express
the change of the velocity deviations of all of the particles at a
$k,l$ collision as a matrix ${\bf M}_{kl}$ acting on the column vector
${\bf{\delta v}}$ whose elements are the velocity deviations,
$\delta\vec{v}_{i}$ for $i=1,...,N$, as ${\bf{\delta v}}'={\bf
M}_{kl}\cdot{\bf{\delta v}}$. Here ${\bf M}_{kl}$ has the value $1$
along the diagonal for the elements corresponding to the $N-2$
particles which are not involved in the $k,l$ collision, and non-zero
elements for the changes in the velocity deviations for particles $k$
and $l$, which can be obtained from Eqs. (13, 14) in terms of the
pre-collision velocity deviations and the times $T_{kl},t_{lk}$. All
other matrix elements of ${\bf M}_{kl}$ are zero.

Thus by neglecting all of the deviations in positions at the instant of
binary collisions
$\delta\vec{r}_{i,0}$, we can easily write an evolution equation for
the velocity deviation vector as the result of a sequence of binary
collisions in the system as
\begin{equation}
{\bf{\delta v}}(t)={\bf M}_{\alpha_1}\cdot{\bf M}_{\alpha_2}\cdots{\bf
M}_{\alpha_{K(t)}}\cdot{\bf{\delta v}}(0).
\label{15}
\end{equation}
In Eq. (\ref{15}) we have used the facts that the dynamics of the
system is a sequence of binary collisions, labeled by the subscripts
$\alpha_j$ with $\alpha_{K(t)}$ the last collision up to time $t$, where only the velocities of
two particles change, separated by free particle motions where none of
the velocities change. For low densities, the product of matrices on
the right hand side of Eq. (\ref{15}) can be thought of as a product
of random $dN \times dN$ matrices, since there are no correlations
between the collisions in the sequences. The random elements of the
sequence are the particles involved in the individual collisions, the
collision parameters of each collision, and the time intervals, for
each particle, between the collisions that it suffers. Under these
circumstances all of the Lyapunov exponents and the KS
entropy of the system can be obtained, in principle, by determining
the eigenvalues of this product of random matrices, using the known
distributions of free times and collision parameters for a dilute gas
in equilibrium. The positive Lyapunov exponents and the KS entropy can
be obtained by using the fact that almost all trajectories in
phase space will separate with time $t$, as $t$ grows large, since the
probability of finding two nearby, but otherwise randomly selected,
trajectories that approach each other for arbitrarily long times, is
vanishingly small. The negative Lyapunov exponents are obtained by
considering the time reversed motion and using the fact that
trajectories that approach each other in the forward time direction
will separate in the time reversed motion. This analysis was applied
to the random, dilute, three dimensional Lorentz gas by Latz, van
Beijeren, and Dorfman \cite{LvBD} who were able to
calculate the positive and negative Lyapunov exponents by analyzing
the products of random matrices similar to those in Eq. (\ref{15})
along these lines. Due to the increased size of the matrices
considered here, the determination of the individual Lyapunov
exponents is still a formidable analytical problem. 
However the determination of the KS entropy is 
relatively elementary as we now show.

Suppose that the eigenvalues of the product of the matrices on the
right hand side of Eq. (\ref{15}) have the form
$\exp(t\lambda_{i})$. Then the logarithm of the determinant of this
matrix product will have the form $t\sum \lambda_{i}$. Since almost
all trajectories will lead to separation, only the positive exponents
will appear in this sum. Further, if we assume that the system is
ergodic then for long times $t$, all possible collision
parameters and free times will appear in the matrices. Thus we can
write
\begin{eqnarray}
h_{KS} &= & \lim_{t\rightarrow\infty}\frac{1}{t}\sum_{j=1}^{j=K(t)}\ln|\det{\bf
M}_{\alpha_j}| \nonumber\\
& = &\frac{N\nu}{2} \left< \ln | \det{\bf M}_{1,2} | \right>.
\label{16}
\end{eqnarray}
where $\nu$ is the average collision frequency per particle, so that
the number of collisions taking place in the gas per unit time is
$N\nu/2$ (each collision involves two particles), and the angular brackets
denote averages over the collision rates and parameters, the free
time distributions, and the velocity distributions for a dilute gas in
equilibrium. We have also used the indices $1,2$ to label the colliding particles in a
typical matrix ${\bf M}$. This matrix has a particularly simple
structure when expressed in terms of the deviations of the center-of-mass
and relative velocities of particles $1,2$ and the velocity
deviations of particles $3,...,N$ which are not affected by the $1,2$
collision,
\begin{equation}
{\bf M}_{1,2}= \left(\begin{array}{ccc}
{\bf 1} & {\bf 0} & {\bf 0} \\
{\bf A} & {\bf B} & {\bf 0} \\
{\bf 0} & {\bf 0} & {\bf 1} \end{array}\right)
\label{17}
\end{equation}
Here we have organized the velocity deviation vector so that ${\bf
\delta v}^{T} = (\delta\vec{V}_{12},
\delta\vec{v}_{21},\delta\vec{v}_{3},...,\delta\vec{v}_{N})^{T}$,
where $T$ denotes a transpose. The four upper-left submatrices are all
$d \times d$ dimensional. The $d \times d$ dimensional submatrices
${\bf A},{\bf B}$, are easily obtained from Eqs. (13, 14)
above. One easily sees that $\det{\bf M}_{1,2} = \det{\bf B}$, which is
simple to calculate. Therefore we find that 
\begin{eqnarray}
h_{KS}/N
& = & \frac{\nu}{2}\left<\ln \left[\,2T_{12}|\vec{v}_{21}|/
        (\sigma\cos\theta)\right]\right>
        \,\, {\rm for}\,\, d=2 \,\, {\rm and} \\
\label{18}
& = & \nu\left<\ln |\left[2T_{12}|\vec{v}_{21}|/\sigma 
    \right] |\right>\,\, {\rm for}\,\, d=3.
\label{19}
\end{eqnarray}
Here $\theta$ is the angle of incidence in a binary collision, with
$-\pi/2 \leq \theta \leq \pi/2$ for two dimensions and $0 \leq \theta
\leq \pi/2$ in three dimensions. Eqs. (18, 19) are the central 
theoretical results of this work. We have obtained explicit
formulae for the KS entropy for a dilute gas of particles with strong
short range forces. We now give the explicit form of the
integrals required for the calculation of $h_{KS}$ for $d=2,3$. For
$d=2$ we obtain 
\begin{eqnarray}
h_{KS}/N =\frac{\nu}{2\pi
J_2}\int d\,\vec{v}_1\phi_0(\vec{v}_1)\int d\,\vec{v}_2\phi_0(\vec{v}_2)
|\vec{v}_{12}|\int_{-\pi/2}^{\pi/2}d\,\theta 
\sigma(\pi-2\theta)/\sigma
\nonumber  \\
\times \nu(v_1)
\nu(v_2)\int_{0}^{\infty}d\tau_1\int_{0}^{\infty}d\,\tau_2
e^{-(\tau_1\nu(v_1)+\tau_2\nu(v_2))}\ln\left[2T_{12}|\vec{v}_{12}|/(\sigma\cos\theta)\right].
\label{20}
\end{eqnarray}
Here $\phi_0(\vec{v})$ is the Maxwell-Boltzmann velocity distribution
function, $\nu(v)$ is the collision frequency for particles with
velocity $\vec{v}$,
$\sigma(\theta)$ is the differential cross section for scattering
under an angle $\theta$ and $J_2$ is a normalization factor obtained by
replacing the logarithmic term in the numerator 
by
unity
and equating the resulting expression to unity.
The corresponding expression for three dimensions is
\begin{eqnarray}
h_{KS}/N = \frac{\nu}{
2\pi
J_3}\int d\,\vec{v}_1 \phi_0(\vec{v}_1) \int\,d
\vec{v}_2 \phi_0(\vec{v}_2) |\vec{v}_{12}| \int_0^{\pi/2} d \theta
\sin \theta \cos \theta 
\sigma(\pi-2\theta)/\sigma^2\int_0^{2\pi} d\phi \nonumber \\
\times \nu(v_1)\nu(\nu(v_2)\int_0^{\infty}\,d \tau_1\int_0^{\infty}\,
d\tau_2 e^{-(\tau_1 \nu(v_1)+\tau_2 \nu(v_2))}
\ln[2T_{12}|\vec{v}_{12}|/\sigma]
\label{20a}
\end{eqnarray}

To evaluate these expressions we need the velocity
dependent collision frequencies for particles with velocities
$\vec{v}_1, \vec{v}_2$, since these frequencies are needed for the
evaluation of the average of $\ln T_{12}$. 
For particles with interactions of finite range expressions for these collision 
frequencies are easily
obtained from elementary kinetic theory, and explicit results can be
given for $h_{KS}$, after some straightforward numerical
integrations. We find that for hard disks ($d=2$) 
\begin{equation}
h_{KS}/N=\frac{\nu}{2}\left[ -\ln(n\sigma^{2}) + 0.209 + O(n)\right]
\label{21}
\end{equation}
where $\nu = [(2\pi^{1/2}n\sigma)/(\beta m)^{1/2}]$ is the
average collision frequency of a dilute gas of hard disks at
temperature $T=(k_B \beta)^{-1}$
, and $k_B$ is Boltzmann's constant. For a gas of hard spheres we find
\begin{equation}
h_{KS}/N = \nu[-\ln(\pi n \sigma^3)+0.562 +O(n)]
\label{22}
\end{equation}
where $\nu = [(4\pi^{1/2}n\sigma^2)/(\beta m)^{1/2}]$.

      To check these results we have carried out extensive 
numerical simulations to compute
the whole set of Lyapunov exponents for equilibrium hard-disk 
and hard-sphere systems. $h_{KS}$ is obtained as 
the sum of all the positive exponents.
In addition to the phase-space trajectory, the ``exact'' molecular-dynamics 
(MD) algorithm
\cite{DPH,DP97} follows the time evolution of a complete set of
infinitesimal vectors $\delta{\bf X}_l$, $l=1,\cdots,2dN$, in tangent space by
using Eqs. (6) and (7) for the streaming between collisions, and 
exact collision maps
analogous to Eq. (\ref{9}) for the discontinuous jumps at a collision. The
tangent vectors are periodically reorthonormalized, and the Lyapunov exponents
are obtained as the time-averaged logarithms of the renormalization factors. 

      The MD-method requires the precise localization of the particle collision
points and becomes inefficient for very low densities. We have therefore
applied also a variant of the direct-simulation Monte Carlo (DSMC) technique 
to the computation of the Lyapunov exponents \cite{DP97}, where the phase-space
dynamics of the system is modeled by appropriate probability distributions
of the collision parameters and of the collision time, and the
tangent-space dynamics remains completely deterministic \cite{DP}.
For very low densities the DSMC-algorithm becomes exact and is 
much more efficient than the MD method. It is expected to fail for
high densities for which correlated collisions become important.

       The simulations, performed with both methods, are for
36 and 64 disks in two dimensions, and for 32 and 108 spheres in three.
The usual periodic boundary
conditions are employed. To present our data
we use reduced units for which the particle diameter $\sigma$, the
kinetic energy per particle, $K/N = m \sum_i\vec{v}_i^2 /2N$, and the
particle mass $m$ are unity. Time is measured in units of 
$(m\sigma^2N/K)^{1/2}$, and the number density $n=N/V$ in units of
$\sigma^{-d}$, where $V$ is the volume of the simulation box. 
$n$ is varied between $10^{-8}$ and $0.1$.  Our choice for the unit of
energy corresponds to a temperature $T$ for which 
$k_B T = K/N = 1$ in two dimensions,
and $k_B T = (2/3) K/N = 2/3$ in three. 
The dependence of
$h_{KS}/(N \nu)$ on the collision frequency $\nu$
is displayed in Figs. \ref{fig_1} and \ref{fig_2} for the respective
two- and three-dimensional systems studied here. 
As indicated by the labels, the diamonds and squares 
denote MD results, whereas the 
plus signs and crosses refer to DSMC data. It is obvious from the 
Figures, that there is very good agreement between the MD and DSMC results 
for $h_{KS}/N \nu$, and that this agreement persists  even
to the largest collision frequencies of interest.
We therefore do not distinguish in the following discussion 
between these two independent sets of simulation data.

    The particle-number dependence of the Kolmogorov-Sinai entropy per
particle is found to be very small for hard disks,
with the points for 36 and
64 particles practically undistinguishable on the 
scale of Fig. \ref{fig_1}.
For hard spheres  there is a noticeable difference between the results for
$N=32$ (dashed) and $N=108$ (dotted) in Fig. \ref{fig_2}. However, it is
expected that the larger of the two systems is already close to the
thermodynamic limit, and that a
comparison of our finite-particle data with
the theoretical prediction, valid for $N \to \infty$, is meaningful.

To facilitate this comparison, we rewrite
Eqs. (\ref{21}) and (\ref{22}) according to
\begin{equation}
    h_{KS}/(N \nu) =  a [-\ln (\nu/\nu_0) + b] + O(\nu^2),
\label{23}
\end{equation}
where
$\nu_0 = 2 (\pi /(m\sigma^2))^{1/2}(K/N)^{1/2}$ for hard disks, and
$\nu_0 = 4 (\pi m \sigma^2)^{-1/2}(2K/3N)^{1/2}$ for hard spheres.
Neglecting the higher-order terms and treating $a,b$ as fit parameters, 
this expression is fitted to the experimental data
for $h_{KS}/(N\nu)$  in the range $10^{-7}<\nu<10^{-2}$.
The results for $a$ and $b$ are summarized in Table \ref{table1}.
The theoretical expectation for $h_{KS}/(N \nu)$ is indicated by the
smooth line in Fig. \ref{fig_1}, $\{a,b\} = \{0.5, 0.209\}$, for disks,
and in Fig. \ref{fig_2}, $\{a,b\} = \{1, 0.562\}$, for spheres. 
            
   From an inspection of Table \ref{table1} we conclude that the theoretically
expected value for
$a$, which determines the slope of the straight lines in the figures, 
is almost perfectly reproduced by the fits. The theoretical values
for the other parameter $b$, however, are not recovered by the 
fitting procedure.
This is reflected in the  offset of the theoretically
expected smooth line with respect to the experimental results in
the Figs. \ref{fig_1} and \ref{fig_2}.  

     To shed some light into the
origin of this disagreement we have numerically evaluated Eq. (\ref{16}) by 
averaging over a molecular-dynamics simulation run. We find perfect agreement 
of this average with the theoretical predictions of Eqs. (\ref{21}) and 
(\ref{22}). 
Moreover, a computer simulation where the terms
$\delta\vec{r}_{k,0}$ and $\delta\vec{r}_{k,0}$ are set equal to zero
for each $k,l$ collision yields results that do agree with Eqs. (22, 23), but 
not with the results where these spatial
displacements are taken into account. Thus, 
there appears to be a
problem with the neglect of the spatial deviations $\delta
\vec{r}_{l,0}-\delta \vec{r}_{k,0}$ in going from Eq. (11) to
Eq. (14). That is, the neglect of
$\delta\vec{r}_{l,0}-\delta\vec{r}_{k,0}$ in Eq. (11) has not yet
been
justified by an analytic calculation where such terms are included. In
fact, 
a more detailed analysis reveals that indeed the inclusion of the neglected
terms yields corrections to the coefficient $b$, which may crudely be estimated to be close to
$\ln 2$.
However these are very
preliminary results 
which must be verified by more extensive
calculations.  
It is worth noting that 
in the case of the two and three
dimensional Lorentz gas 
the neglect of spatial deviations in the equivalent of Eq.~\ref{14} can be
fully justified
and there the theoretical results for the
analogs of the coefficients $a$ and $b$ agree very well with the
simulation results \cite{vBD,vBDCPD,LvBD,DP,DPII}.
Finally notice that the assumptions made do not affect
the theoretical values of the leading term
involving only the parameter $a$. 
Presently we are working both on calculations of the terms neglected above
and on a more systematic
approach 
using methods based upon
the BBGKY hierarchy equations, which will provide a
systematic density expansion
of $h_{KS}$. 

In conclusion, we have obtained the first theoretical values for the
leading term in the KS entropy per particle for dilute gases with
short range forces, both for two and three dimensional systems. The
comparison of these predictions with results of computer simulations
are excellent, but 
we do not yet have a final analytic expression for
the first order
corrections to these results. 
The computer simulations, however, do provide accurate values for these
corrections, to which eventually analytic results can be compared. 
There seems to be little doubt that
our results can be extended to dense gases and to non-equilibrium
states of interest for the escape-rate or the Gaussian thermostat
approaches for relating transport coefficients to the chaotic
properties of fluid systems.

ACKNOWLEDGEMENTS

The authors are pleased to acknowledge the hospitality of the Erwin
Schr\"{o}dinger Institut of the Universit\"{a}t Wien where this
work was initiated in December, 1996. 
We would like to thank Dr. Arnulf Latz and
Mr. Ramses van Zon for helpful comments on this and related subjects;
Dr. Charles Ferguson for helpful comments and assistance with
numerical integrations; and the National Science Foundation for
support under grant NSF-PHY-96-00428. HAP and CD gratefully acknowledge
support from the Fonds zur F\"orderung der wissenschaftlichen Forschung,
under Grants P09677-PHY, P11428-PHY, and J01302-PHY and HvB acknowledges support
by FOM, SMC and by the NWO Priority Program Non-Linear
Systems, which are financially supported by the "Nederlandse Organisatie voor Wetenschappelijk Onderzoek (NWO)".

\newpage

\begin{table}
\caption{ Fit parameters for the fit of Eq. (23) 
to the experimental $h_{KS}/(N \nu)$,
for a system containing $N$ hard disks, $(d=2)$,
or spheres, $(d=3)$. The fitting range is $10^{-7}< \nu <10^{-2}$,
where $\nu$ is the collision frequency in the reduced units
introduced in the main text. MD and DSMC refer to the numerical
method used for the computation. The quoted errors for the
fit parameters $a$ and $b$ are standard deviations.}

\label{table1}
\begin{tabular}{crlcc}
$d$ & $N$  &  Method    &    a      &        b            \\
\tableline
2 & 36 & MD  & 0.499  $\pm$  0.001 & 1.339  $\pm$  0.003  \\
2 & 36 & DSMC& 0.500  $\pm$  0.001 & 1.326  $\pm$  0.006  \\
2 & 64 & MD  & 0.499  $\pm$  0.001 & 1.366  $\pm$  0.005  \\
2 & 64 & DSMC& 0.500  $\pm$  0.001 & 1.358  $\pm$  0.009  \\
\tableline
2 & $\infty$ & Theory & 0.5        &  0.209  \\
\tableline
3 & 32 & MD  & 0.998  $\pm$  0.001 & 1.42   $\pm$  0.01   \\
3 & 32 & DSMC& 1.000  $\pm$  0.001 & 1.38   $\pm$  0.01   \\
3 & 108& MD  & 1.03   $\pm$  0.03  & 1.34   $\pm$  0.29   \\
3 & 108& DSMC& 1.036  $\pm$  0.002 & 1.34   $\pm$  0.02   \\
\tableline
3 & $\infty$ & Theory & 1.0        &  0.562 \\
\end{tabular}
\end{table}

\section*{Figure Captions}
\begin{enumerate}
\item \label{fig_1}  $h_{KS}/(N \nu) $, as
a function of the collision frequency, for planar systems containing 
36 and 64 hard disks.
$\nu$ is measured in the reduced units introduced
in the main text. The points are labeled according to the computational
method, MD or DSMC, the dimension, $d=2$, and the number of particles $N$.
The smooth line is the theoretical prediction according to
Eq. (\ref{21}), and the (almost undistinguishable) dashed and dotted 
lines are fits  of Eq. (\ref{23}) to the experimental points
for $N=36$ and $N=64$, respectively. The fitting range is
$10^{-7}< \nu <10^{-2}$. The fit parameters $a$ and $b$ are
listed in Table \ref{table1}.

\item \label{fig_2}  $h_{KS}/(N \nu) $, as
a function of the collision frequency, for three-dimensional systems containing 
32 and 108 hard spheres.
$\nu$ is measured in the reduced units introduced
in the main text. The points are labeled according to the computational
method, MD or DSMC, the dimension, $d=3$, and the number of particles $N$.
The smooth line is the theoretical prediction according to
Eq. (\ref{22}), and the dashed and dotted 
lines are fits  of Eq. (\ref{23}) to the experimental points
for $N=32$ and $N=108$, respectively. The fitting range is
$10^{-7}< \nu <10^{-2}$. The  fit parameters $a$ and $b$ are
listed in Table \ref{table1}.

\end{enumerate}
\end{document}